\documentclass[twocolumn,longbibliography]{revtex4-1}

\usepackage{amssymb}

\usepackage{tikz}
\usetikzlibrary{decorations.markings}

\usepackage[pdfpagelabels,pdftex,bookmarks,breaklinks]{hyperref}
\hypersetup{colorlinks,citecolor=blue,urlcolor=blue,linkcolor=black}

\newtheorem{theorem}{Theorem}
\newtheorem{lemma}{Lemma}
\newtheorem{problem}{Problem}

\begin{document}

\tikzset{->-/.style={decoration={
  markings,
  mark=at position #1 with {\arrow{>}}},postaction={decorate}}
  }

\title{A Quantum Approach to the Unique Sink Orientation Problem}
\author{Dave Bacon}
\affiliation{Google, Seattle, WA 98103 USA}
\email{dabacon@google.com}
\begin{abstract}
We consider quantum algorithms for the unique sink orientation problem on cubes.  This problem is widely considered to be of intermediate computational complexity.  This is because there no known polynomial algorithm (classical or quantum) from the problem and yet it arrises as part of a series of problems for which it being intractable would imply complexity theoretic collapses.   We give a reduction which proves that if one can efficiently evaluate the $k$th power of the unique sink orientation outmap, then there exists a polynomial time quantum algorithm for the unique sink orientation problem on cubes.    
\end{abstract}
\maketitle

In this paper we are concerned with finding an efficient quantum algorithm for a problem that admits no known classical polynomial time algorithm despite considerable effort, the unique sink orientation problem on cubes.  In this problem one is given a directed graph on a hypercube such that every face (subcube) of the hypercube admits a unique sink vertex, and the goal is to find the global unique sink of the entire cube.  Access to information about an instance of this problem is via an oracle which takes as inputs one of the vertices of the cube and outputs a list of the directions of the outgoing edges of that vertex.  The best (classical) algorithm for this problem queries the oracle $O((1.467\dots)^n)$ times where $n$ is the dimension of the hypercube~\cite{szabo2001unique}, or, if the directed graph has no cycles $O(\exp(n^{1/2}))$ times~\cite{gartner1998combinatorial, gartner2002random}.  We are not able to obtain a polynomial time quantum algorithm for this problem, but we are able to show that if one could efficiently calculate the $k$th power of the oracle in this problem (defined below), then there exists an efficient quantum algorithm.  Our path to this is through period finding, the key ingredient to Shor's efficient quantum algorithm for factoring~\cite{shor1994algorithms, shor1999polynomial}.  The main difference in our failure versus Shor's success is that in Shor's algorithm there are efficient algorithms to calculate the $k$th power of a number mod $N$ for $k$ exponentially large (modular exponentiation), whereas for the unique sink orientation problem on cubes we do not yet have such a procedure.  Our reduction opens a new approach towards obtaining an efficient quantum algorithm for the unique sink orientation problem on cubes.

\section{Background and Motivation}

The unique sink orientation problem on cubes (hereafter abbreviated as the \textsc{USO} problem) arises as a fundamental problem in a variety of optimization problems. 
The original application of \textsc{USO} problem comes from the observation that a polynomial time algorithm for the \textsc{USO} problem would yield a polynomial time algorithm for a class of linear complementarity problems that have no known polynomial time algorithm (those arising from P-matrices)~\cite{cottle1992linear}.  Another application is to linear programming.  Recall that a numerical problem is strongly polynomial if assuming unit cost for arithmetic on the involved numerical quantities, the algorithm takes takes a polynomial amount of time in the number of numerical constants. While there are weakly polynomial time algorithms for linear programming~\cite{khachiyan1980polynomial}, there is no known strongly polynomial time algorithm.  A polynomial time algorithm for the \textsc{USO} problem, however, would imply a strongly polynomial time algorithm for linear programming~\cite{gartner2006linear}.  A variety of problems in the theory of games would also admit polynomial time solution~\cite{gartner2005simple, svensson2005mean, vorobyov2008cyclic} if there is a polynomial time algorithm algorithm for the \textsc{USO} problem.  Finally, finding the smallest ball enclosing a set of ball would also admit a polynomial time solution if there was a polynomial time algorithm for the \textsc{USO} problem~\cite{fischer2004smallest}.

The \textsc{USO} problem is a promise problem: we are promised that the faces of the hypercube graph all have a unique sink. This promise is itself not known to be polynomial time verifiable~\cite{gartner2015complexity}.  Therefore (a decision) version of this problem does not fit nicely into a discussion of computational complexity.  However there is strong evidence that the problems which motivate investing \textsc{USO} are not computationally intractable.  For example, the P-matrix linear complementarity problem is not \textsc{NP-hard} unless \textsc{NP}=\textsc{coNP}~\cite{megiddo1988note}.  Because of this the \textsc{USO} problem is widely considered as a candidate for either having a polynomial time algorithm, or being of intermediate computational complexity, similar to the status of the factoring problem.

A variety of classical algorithms exists for the \textsc{USO} problem.  Most of these are known to have instances on which they take an exponentially long time.  For example the algorithm of randomly following one outgoing edge of a vertex, {\rm RANDOM EDGE}, can be shown to take $\Omega(\exp(n^{1/3}))$ time~\cite{matousek2004random}.  Similarly the algorithm of choosing a random facet, and then recursing, can also be shown to take $\Omega(\exp(n^{1/2}))$ time~\cite{gartner2002random}.   These results all give instances upon which the given algorithms fail to work in polynomial time.  In a similar vein we will show below that a naive classical version of our quantum algorithm that uses the standard oracle to solve the \textsc{USO} problem requires exponential time.

Despite the \textsc{USO} problem being of intermediate computational complexity the only quantum computation work for this problem has been on the problem of recognizing whether an orientation is a unique sink orientation or not~\cite{bosshard2015classical}.  We note in passing that a naive application of Grover's algorithm~\cite{grover1996fast} to the general \textsc{USO} problem yields an $O(\sqrt{2^n})=O((1.414\ldots)^n)$ quantum algorithm, which just marginally beats the best known classical algorithm which is $O((1.467\dots)^n)$~\cite{szabo2001unique}.

\section{Unique Sink Orientation Problem}

Here we introduce the \textsc{USO} problem and present some prior results that will be useful.  An excellent well-written introduction to this problem is the Ph.D. thesis of I. A. Schurr~\cite{schurr2004unique}.  We begin with an information description of the problem and then proceed to a more formal specification which is useful for stating basic useful properties of the \textsc{USO} problem.

An instance of the \textsc{USO} problem is an orientation of a boolean (combinatorial) hypercube which special properties.  Recall that the boolean hypercube is the graph in which vertices of the graph are boolean strings and there is an edge between two vertices if these vertices differ in only one character of the string.  Thus $0010$ is a vertex and $0011$ is one of its neighbors while $1011$ is not one of its neighbors.  An orientation is a specification of a direction for each of the edges in the graph.  There are $2^{2^n}$ orientations.  In the \textsc{USO} problem, however, a constraint is imposed on the orientation.  Recall that an a vertex in a graph with directed edges is a sink if all the edges adjacent to a vertex are incoming.  In the \textsc{USO} problem, the orientation is required to satisfy the condition that if one restricts one attention a subcube, only one of these vertices is a sink under this restriction.  By restrict to a subcube we mean consider a $k \leq n$ hypercube and only examine the orientation of edges between the vertices of this subcube.

Next we introduce the notation for the combinatorial hypercube we will use in this paper.  This notation is that used by the literature about USOs and is used here to help provide a bridge into that literature and also to make the proofs that appear in Sec.~\ref{sec:properties} cleaner.  Let $[n]=\{1,2,\dots,n\}$.  We will label the vertices of the $n$-dimensional hypercube by the subsets of $[n]$.  That is vertices correspond to the power set of $[n]$: $V=2^{[n]} :=\{ v \subset [n]\}$.  If one prefers to think about the hypercube vertices as boolean strings, then one can consider a subset of $[n]$ as corresponding to the elements in an $n$ bit string that are $1$, while those not in the subset are $0$.   The symmetric difference of two sets $u$ and $v$ is the set of elements that are in $u$ and $v$ but not both, $u \oplus v := (u \cup v) \setminus  (u \cap v)$.  Given this definition the edges of the $n$-dimensional hypercube are then vertices whose symmetric difference has size $1$.  $E = \{ (u,v)| u,v \subseteq [n], |u \oplus v|=1\}$.  This the version of the condition that an edge in a hypercube exists between boolean strings if they differ in exactly one position.

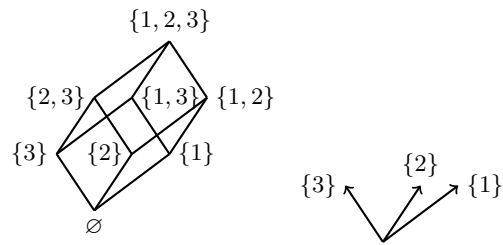
\begin{figure}[b]
\begin{minipage}{0.4\textwidth}
\begin{tikzpicture}
 \draw[thick,-=.5] (0,0) to (-0.5,0.75);
 \draw[thick,-=.5] (0,0) to (1,0.75);
 \draw[thick,-=.5] (1,0.75) to (0.5,1.5);
 \draw[thick,-=.5] (-0.5,0.75) to (0.5,1.5);
 
 \draw[thick,-=.5] (0.5,0.75) to (0,1.5);
 \draw[thick,-=.5] (0.5,0.75) to (1.5,1.5);
 \draw[thick,-=.5] (1.5,1.5) to (1,2.25);
 \draw[thick,-=.5] (0,1.5) to (1,2.25);

 \draw[thick,-=.5] (0,0) to (0.5,0.75);
 \draw[thick,-=.5] (-0.5,0.75) to (0,1.5);
 \draw[thick,-=.5] (1,0.75) to (1.5,1.5);
 \draw[thick,-=.5] (1,2.25) to (0.5,1.5);

\draw[below,font=\small] (0,0) node {$\varnothing$};
\draw[left,font=\small] (-0.5,0.75) node {$\{3\}$};
\draw[right,font=\small] (1,0.75) node {$\{1\}$};
\draw[right,font=\small] (0.5,1.5) node {$\{1,3\}$};

\draw[left,font=\small] (0.5,0.75) node {$\{2\}$};
\draw[left,font=\small] (0,1.5) node {$\{2,3\}$};
\draw[right,font=\small] (1.5,1.5) node {$\{1,2\}$};
\draw[above,font=\small] (1,2.25) node {$\{1,2,3\}$};
\end{tikzpicture}
\begin{tikzpicture}
 \draw[thick,->=.5] (0,0) to (-0.5,0.75);
 \draw[thick,->=.5] (0,0) to (1,0.75);
 \draw[thick,->=.5] (0,0) to (0.5,0.75);
\draw[right,font=\small] (1,0.75) node {$\{1\}$};
\draw[above,font=\small] (0.5,0.75) node {$\{2\}$};
\draw[left,font=\small] (-0.5,0.75) node {$\{3\}$};
\end{tikzpicture}
\end{minipage}
\caption{Labeling of the vertices of a hypercube for $n=3$ on the left and the different directions on the right.} \label{fig:hypercube}
\end{figure}

More generally, for sets $u \subseteq v$ define the interval of these sets as $[u:v] := \{ w | u \subseteq w \subseteq v\}$.  Then the cube spanned by these sets ${\mathcal C}^{[u:v]}$ is defined as the graph with vertex set
\begin{equation}
V({\mathcal C}^{[u:v]}) := [u:v]
\end{equation}
and edge set
\begin{equation}
E({\mathcal C}^{[u:v]}) := \{(s,t) | s,t \in [u:v], |s \oplus t | = 1\}.
\end{equation}
A useful shorthand is ${\mathcal C}^u = {\mathcal C}^{[\varnothing:u]}$, in which case the $n$-dimensional hypercube is ${\mathcal C}^{[n]}$.  Intervals are useful for defining subcubes. The cube ${\mathcal C}^{[u,v]}$ is a subcube of the cube ${\mathcal C}^{[u^\prime:v^\prime]}$ iff $[u,v] \subseteq [u^\prime:v^\prime]$.  Subcubes of a cube are often called faces and a face of dimension one less than the cube is called a facet.  If one prefers to work over the boolean string representation of a hypercube, then the vertices of a hypercube are those where a fixed set of character positions are fixed, and other are allowed to vary.  The elements that are in $v$ but not in $u$ of a subcbue ${\mathcal C}^{[u,v]}$ are exactly the locations where the characters are allowed to vary.

Edges of the $n$-dimensional hypercube are labeled in a natural way by their direction label.  That is for the edge $e=(u,v)$ there is a unique $\lambda$ such that $\{ \lambda  \}= u \oplus v$.  The set of all labels of a cube is called the carrier
\begin{equation}
{\rm carr}~{\mathcal C}^{[u:v]}:=v \setminus u.
\end{equation}
For a given direction $\lambda$ we can split the cube into two subcubes, called the $\lambda$-facets.  In particular for the cube ${\mathcal C}^{[n]}$, the lower $\lambda$-facet is ${\mathcal C}^{[n] \setminus \{\lambda\}}$ and the upper $\lambda$-facet is ${\mathcal C}^{[\{\lambda\}:[n]]}$.  Less formally the condition is that when one considers only the orientations of the edges between the vertices in a subcube, this orientation has a unique sink vertex.  See Fig.~\ref{fig:hypercube} for this labeling scheme and direction in the case of $n=3$.

An orientation $\phi$ of a graph $G=(V,E)$ is a map from the edges $E$ to the vertices $V$ such that $\phi((v,w))=v$ or $\phi((v,w))=w$ for all $(v,w) \in E$.  This maps edges to their corresponding sink vertex.  Thus if $(v,w) \in E$ and $\phi((v,w)) = v$ then the orientation has the edge $E$ directed from $w$ to $v$.  In this case $w$ is the source and $v$ is the sink. Given this notion for a particular vertex we can partition edges that contain the vertex into those for which the vertex is the sink (incoming edges) and those for which it is the source (outgoing edges).  

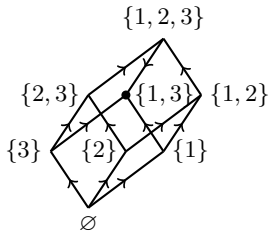
\begin{figure}[t]
\begin{tikzpicture}
 \draw[thick,->-=.5] (0,0) to (-0.5,0.75);
 \draw[thick,->-=.5] (0,0) to (1,0.75);
 \draw[thick,->-=.5] (1,0.75) to (0.5,1.5);
 \draw[thick,->-=.5] (-0.5,0.75) to (0.5,1.5);
 
 \draw[thick,->-=.5] (0.5,0.75) to (0,1.5);
 \draw[thick,->-=.5] (0.5,0.75) to (1.5,1.5);
 \draw[thick,->-=.5] (1.5,1.5) to (1,2.25);
 \draw[thick,->-=.5] (0,1.5) to (1,2.25);

 \draw[thick,->-=.5] (0,0) to (0.5,0.75);
 \draw[thick,->-=.5] (-0.5,0.75) to (0,1.5);
 \draw[thick,->-=.5] (1,0.75) to (1.5,1.5);
 \draw[thick,->-=.5] (1,2.25) to (0.5,1.5);
\fill[black] (0.5,1.5) circle (0.06);

\draw[below,font=\small] (0,0) node {$\varnothing$};
\draw[left,font=\small] (-0.5,0.75) node {$\{3\}$};
\draw[right,font=\small] (1,0.75) node {$\{1\}$};
\draw[right,font=\small] (0.5,1.5) node {$\{1,3\}$};

\draw[left,font=\small] (0.5,0.75) node {$\{2\}$};
\draw[left,font=\small] (0,1.5) node {$\{2,3\}$};
\draw[right,font=\small] (1.5,1.5) node {$\{1,2\}$};
\draw[above,font=\small] (1,2.25) node {$\{1,2,3\}$};
\end{tikzpicture}
\caption{Example of a \textsc{USO}.  The unique sink is vertex $\{1,3\}$ and the unique source is $\varphi$.
Note, for instance that this is a \textsc{USO} since all of its faces considered alone have a unique sink vertex.  For example the ${\mathcal C}^{[\varnothing:\{1,2\}]}$ cube has the unique sink of $\{1,2\}$.} \label{fig:uso}
\end{figure}

Given an orientation $\phi$ of a cube ${\mathcal C}$, we define the outmap $s$ of $\phi$ as the map that assigns to every vertex the labels of the outgoing edges from that vertex:
\begin{equation}
s :  V({\mathcal C}) \rightarrow 2^{{\rm carr}~{\mathcal C}}
\end{equation}
We can now define a \textsc{USO} of a cube ${\mathcal C}$.  A \textsc{USO} of a cube ${\mathcal C}$ is an orientation $\phi$ such that every subcube has a unique sink.  That is an orientation $\phi$ with outmap $s$ is a \textsc{USO} iff 
\begin{equation}
\forall u,v~C^{[u,v]}~{\rm is~a~unique~sink~orientation.}
\end{equation}
where a subcube $C^{[u,v]}$ is a unique sink orientation if there exists a unique vertex $w$ such that the outmap does not point along any of the outgoing directions in the carrier space for a subcube, ${\rm carr}~{\mathcal C}^{[u:v]}$,
\begin{equation}
\exists~{\rm unique}~w~{\rm such~that~} \forall \lambda \in {\rm carr}~{\mathcal C}^{[u:v]}, s(w) \notin \lambda
\end{equation}
See Fig.~\ref{fig:uso} for a picture of a \textsc{USO} for $n=3$.

The central problem we would like to solve can now be introduced.
\begin{problem}[\textsc{USO}] \label{pr:uso}
Given a \textsc{USO} $\phi$ on a cube ${\mathcal C}^{[n]}$ and a subcube ${\mathcal C}^{[u:v]}$, determine by querying the outmap $s$ of $\phi$ whether the unique sink of ${\mathcal C}^{[n]}$ lies in ${\mathcal C}^{[u:v]}$.
\end{problem}
Note that this is a decision problem having a yes/no answer for each $\phi$ and subcube ${\mathcal C}^{[u:v]}$.  However it can be used to efficiently solve the search version of the problem, finding the unique sink vertex of the entire cube.  To do this one simply uses the decision version on facets of the cube, determining whether the unique sink is in one of two facets, and then recursively applying this procedure to the facet with the unique sink.

\section{Properties of \textsc{USO}s} \label{sec:properties}

Here we recall certain properties of \textsc{USO}s on cubes that we will use.  We present short proofs of these results for completeness.

If $\phi$ is an orientation with corresponding outmap $s$ for a cube ${\mathcal C}$, then the orientation $\phi^\prime$ in which for every $\lambda \in \Lambda \subseteq {\rm carr}~{\mathcal C}$ the $\lambda$-edge of $\phi$ are reversed has an outmap of $s^\prime(v) = \Lambda\oplus s(v)$.  If the original orientation $\phi$ is a \textsc{USO}, then, according to the following Lemma, the orientation $\phi^\prime$ is also a \textsc{USO}:
\begin{lemma} \label{lemma:flip}
\cite{schurr2004unique}~Given a unique sink outmap $s$ of a cube ${\mathcal C}^{[n]}$, and a set of directions, $\Lambda \subseteq {\rm carr}~{\mathcal C}^{[n]}$, the map defined by $s^\prime(v) = \Lambda \oplus s(v)$ is the outmap of a \textsc{USO}.    
\end{lemma}

PROOF. We will first show this is true for $\Lambda = \{ \lambda \}$.  We need to show that the new outmap $s^\prime$ corresponds to an orientation $\phi^\prime$ that has the \textsc{USO} property.  Let ${\mathcal C}_u$ and ${\mathcal C}_l$ denote the upper and lower $\lambda$-facets of ${\mathcal C}^{[n]}$ respectively.  Consider any subcube ${\mathcal C}$ of ${\mathcal C}^{[n]}$.  If ${\mathcal C}$ is entirely within a $\lambda$-facet, then the orientation obtained by reversing all edges along the $\lambda$ direction does not change the orientation on any edges of ${\mathcal C}$.  In this case since $\phi$ has a unique sink on ${\mathcal C}$, so does $\phi^\prime$ on ${\mathcal C}$.  On the other hand, if ${\mathcal C}$ spans $\lambda$-facets, then we can partition ${\mathcal C}$ into sub cubes ${\mathcal D}_u = {\mathcal C} \cup {\mathcal C}_u$ and ${\mathcal D}_d = {\mathcal C} \cup {\mathcal C}_d$.  Over the orientation $\phi$ suppose that these two subcubes have unique sinks $o_u$ and $o_d$.  Assume without loss of generality that the unique sink of $\phi$ over ${\mathcal C}$ is $o_u$.  Flipping all of the edges along $\lambda$ means that $o_u$ is no longer a unique sink (since it's $\lambda$ edge now points away from it).  However $o_d$ must now be a unique sink of ${\mathcal C}$ since prior to flipping the $\lambda$ edge the only thing keeping $o_d$ from being a unique sink of ${\mathcal C}$ was the $\lambda$ edge.  All other vertices cannot be unique sinks of ${\mathcal C}$ since they are not unique sinks in ${\mathcal D}_u$ or ${\mathcal D}_d$ and none of the edges in those sub cubes are flipped.  Hence $\phi^\prime$ is the outmap of a \textsc{USO} for $\Lambda = \{ \lambda \}$.  For the general case of $\Lambda=\{\lambda_1,\dots,\lambda_k\}$ note that it follows from the application of the just proven case $k$ times since $s^\prime(v) = \Lambda \oplus s(v) = \{\lambda_1\} \oplus \cdots \oplus \{\lambda_k\} \oplus s(v)$.

The use of the above lemma is that it implies an important property of all outmaps of a \textsc{USO} of a cube.    
\begin{lemma}
\cite{hoke1988complete}~The outmap of a \textsc{USO} of a cube ${\mathcal C}$ is a bijection.
\end{lemma}

PROOF. Given a \textsc{USO} $\phi$ of a cube ${\mathcal C}$.  Suppose that there exists two vertices, $u \neq v$ which each have the same image under the outmap $s$, $s(v)=s(u)=t$.  Then consider the orientation $\phi^\prime$ obtained by flipping all of the edges along the $t$ direction.  This has outmap $s^\prime(v)= t \oplus v$.  Via Lemma~\ref{lemma:flip} this new orientation is a \textsc{USO}.  However $s^\prime(u)=s^\prime(v) = \varnothing$, which is a contradiction.  Hence there are not two vertices which have the same image under the outmap $s$.

\section{Quantum Approach}

In the last section we have seen that the outmap of a \textsc{USO} is a bijection from vertices to the power set of the carrier space of the cube.  Vertices are labeled by elements of $2^{[n]}$ and elements of the carrier space are also labeled by $2^{[n]}$.  Viewed in this manner we can map the carrier space back to the vertices, and hence we can view the outmap as a permutation on $2^{[n]}$.  Given this view we can then define the $k$th power of this map
\begin{equation}
s^k := \underbrace{s \circ s \circ \cdots \circ s}_{k~{\rm times}}
\end{equation}
i.e. the permutation $s$ applied $k$ times.

Consider the sequence 
\begin{equation}
\varnothing, s(\varnothing), s^2(\varnothing), \dots
\end{equation}
Because $s$ is a bijection this sequence is periodic with a period at most $2^n$.  That is there exists a minimal $l>0$ such that $s^k(\varnothing) = s^{k+l}(\varnothing)$.

Suppose that we could determine the minimal $l \neq 0$ such that $s^k(\varnothing) = s^{k+l}(\varnothing)$.  Then $s^l(\varnothing)=\varnothing$ and hence $s (s^{l-1}(\varnothing))=\varnothing$, or in other words $s^{l-1}(\varnothing)$ is the unique sink of the cube.  In other words determining the period of $s$, and then evaluating $s$ raised to that power minus one, yields the unique sink.  The problem of finding the period of a function is one in which quantum computers offer exponential advantage over classical computers, and this is the basis of Shor's algorithm for factoring~\cite{shor1994algorithms, shor1999polynomial}.

In particular this leads us to our main result.  Let $\phi$ be a \textsc{USO} with outmap $s$.  Let ${\mathcal H}=\mathbb{C}_2^n \otimes \mathbb{C}_2^n \otimes \mathbb{C}_2^n$ and label the basis of these states by $|k,u,v\rangle$ with $k \in \{0,\dots,2^n-1\}$ and $u,v \in 2^{[n]}$.  We say that a unitary $U$ calculates the $k$th power of an outmap if acts on this basis as
\begin{equation}
U|k,u,v\rangle = |k,u,v \oplus s^k(u)\rangle.
\end{equation}

\begin{theorem}
Given a unitary $U$ which can calculate the $k$th power of the outmap permutation of a \textsc{USO}, there exists a polynomial sized quantum algorithm that queries this oracle $O(1)$ times and identifies the unique sink with constant probability with a circuit of size $O(poly(n))$.
\end{theorem}

PROOF. Straightforward application of Shor's period finding algorithm to the given oracle~\cite{shor1994algorithms, shor1999polynomial}, followed by using the oracle once to calculate $s^{l-1}(\varnothing)$.

\section{Exponentially long periods}

In the previous section we showed that if one could efficiently determine the period of an outmap $s$ of a \textsc{USO} of a cube and one can evaluate the outmap at this period minus one, then one can efficiently solve the \textsc{USO} problem.  A question which arises is whether or not this an efficient classical algorithm.  A naive classical use of this period finding method would be try to calculate$s(\varnothing), s^2(\varnothing), \dots$ until the sequence repeats.  Here we show that this approach fails because there are \textsc{USO}s where this sequence is exponentially long.

We will explicitly show an example of such a \textsc{USO}.  Given two \textsc{USO}s of the same dimension $n$, $\phi_1$ and $\phi_2$, one can produce a new \textsc{USO} of dimension $n+1$ by using $\phi_1$ as a lower $n+1$ face and $\phi_2$ as the upper $n+1$ face and then directing all of the $n+1$ edges either from the lower to the upper face, or vice versa.  Label these two cases $\phi_1 \uparrow \phi_2$ and $\phi_1 \downarrow \phi_2$, respectively.  It is clear that the new orientation constructed in this manner is a \textsc{USO}, since every face either lies entirely in $\phi_1$ or $\phi_2$ (and so is a \textsc{USO}), or the face spans $\phi_1$ and $\phi_2$ and hence each component in the upper and lower $n+1$ face has a unique sink so the global \textsc{USO} exists and is uniquely the one in which the new $n+1$ orientation points.

Let $\psi_1$ be the $1$ dimensional orientation in which the single edge points from $\varnothing$ to $\{1\}$.  Further define the uniform orientation $u_{n,a}$ as the $n$ dimensional orientation whose outmap is $s(v)=v \oplus a$.  Recursively define the orientation
\begin{equation}
\psi_{n+1} = \psi_n \downarrow u_{n-1,\varnothing}
\end{equation}
This orientation has $\psi_n$ as it's lower $n+1$ facet, the uniform orientation towards $\varnothing$ in its upper $n+1$ facet, and $n+1$ edges from the upper to the lower $n+1$ facet.

\begin{lemma}
Define $P(\phi)$ as the period of the outmap of $\phi$, starting at $\varnothing$.  Then
\begin{equation} 
P(\psi_{n+1}) = 2{P(\psi_n)}
\end{equation}
and hence since $P(\psi_1)=2$, $P(\psi_n)=2^n$.
\end{lemma}
PROOF. Let $s_n$ denote the outmap of $\psi_n$.  The outmap of $\psi_{n+1}$ can be written as
\begin{equation}
s_{n+1}(v) = 
\left\{
\begin{array}{l} 
	\{n+1\} \cup s_{n}(v)~{\rm if}~{n+1 \notin v} \\ 
	v \setminus \{n+1\} ~{\rm otherwise} 
 \end{array} 
\right.  \label{eq:firstflip}
\end{equation}
From this it follows that the sequence $s_{n+1}(\varnothing), s_{n+1}^2(\varnothing), s_{n+1}^3(\varnothing), s_{n+1}^4(\varnothing), \dots$ is equal to $s_{n}(\varnothing) \cup \{n+1\}, s_{n}(\varnothing), s_{n}^2(\varnothing) \cup \{n+1\}, s_{n}^2(\varnothing) \dots$.  Since $s_n$ does not act on the $n+1$ element, the claim follows directly.

Because there exist \textsc{USO}s like $\phi_n$ that have exponentially long periods the naive classical algorithm for finding the period fails.   Note that this does not mean that a classical approach based upon identifying the period will fail, only that the naive algorithm would take exponentially long.  Perhaps \textsc{USO}s have structure which will would allow for an efficient classical algorithm based upon finding the period of the outmap.

\section{The Missing Piece}

We have shown that the ability to efficiently calculate the $k$th power of a \textsc{USO}s outmap would lead to an efficient quantum algorithm for the problem and that the same is not necessarily true for a naive classical algorithm.  In Shor's algorithm the step we are replacing with the $k$th power of a \textsc{USO} is the evaluation of $r^x~{\rm mod}~N$.  This can be done by modular exponentiation (using the trick of repeated squaring).  We have been unable to find an efficient way to calculate this $k$th power.  In general it would seem that such a procedure would need to rely on properties of \textsc{USO}s.  Note that in general there is no procedure that efficiently calculates the power of an arbitrary unitary gate.  This follows from \cite{berry2007efficient} where it was shown that evaluating $\exp(-iHt)$ scales at list linearly in $t$, though this does not preclude such powering for known sets of unitaries (see also \cite{soleimanifar2016no,atia2016fast}). There are a large classes of \textsc{USO}s which have structure because they arise from different algorithmic problems~\cite{foniok2014counting}.  For example those arising in the P-matrix linear complementarity problem are known to be a restricted class that is rather small relative to all \textsc{USO}s and yet efficiently solving the \textsc{USO} arising from these problem would constitute a breakthrough.  Focusing on these instances is suggested as an important future direction.

\section{Conclusion}

We have shown that the ability to efficiently calculate the $k$th power of a \textsc{USO}s outmap would imply that there is an efficient quantum algorithm for the \textsc{USO} problem.  This approaches uses a key component that is thought to be important for quantum speedups~\cite{childs2010quantum}, period finding.  A variety of open problems remain, even beyond the obvious of trying to efficiently power the outmap.  One important question is whether there are classes of \textsc{USO}s for which there is a quantum speedup.  For example, \textsc{USO}s that are decomposable are known have query complexity $\Theta(n)$~\cite{schurr2002finding, schurr2004unique}, is there a constant query quantum algorithm for these cases?  Another interesting question is whether the query complexity of the \textsc{USO} is  polynomial in an information theoretic sense.

\bibliography{uso_arxiv_v2}{}
\end{document}